\documentclass[sigconf]{acmart}
\usepackage{balance}
\usepackage{graphics}
\usepackage{color}
\usepackage{booktabs}
\usepackage{subcaption}
\usepackage{todonotes}
\usepackage{arydshln}
\usepackage{colortbl}

\AtBeginDocument{%
  \providecommand\BibTeX{{%
    \normalfont B\kern-0.5em{\scshape i\kern-0.25em b}\kern-0.8em\TeX}}}

\copyrightyear{2021} 
\acmYear{2021} 
\setcopyright{acmlicensed}
\acmConference[CHIIR '21] {Proceedings of the 2021 ACM SIGIR Conference on Human Information Interaction and Retrieval}{March 14--19, 2021}{Canberra, Australia}
\acmBooktitle{Proceedings of the 2021 ACM SIGIR Conference on Human Information Interaction and Retrieval (CHIIR '21), March 14--19, 2021, Canberra, Australia}
\acmPrice{15.00}
\acmISBN{978-1-4503-8055-3/21/03} 
\acmDOI{10.1145/3406522.3446038}

\begin{document}
\fancyhead{}

    \definecolor{orange}{RGB}{255,90,20}
\definecolor{blue}{RGB}{20,90,255}
\newcommand{\TODO}[1]{\textbf{\textcolor{orange}{#1}}}

\title{Impact of Response Latency on User Behaviour in\\ Mobile Web Search}

\author{Ioannis Arapakis}
\affiliation{
  \institution{Telefonica Research}
  \country{Spain}
}
\email{ioannis.arapakis@telefonica.com}

\author{Souneil Park}
\affiliation{
  \institution{Telefonica Research}
  \country{Spain}
}
\email{souneil.park@telefonica.com}

\author{Martin Pielot}
\affiliation{
  \institution{Google}
  \country{Germany}
}
\email{mpielot@google.com}

\begin{abstract}
Traditionally, the efficiency and effectiveness of search systems have both been of great interest to the information retrieval community. However, an in-depth analysis of the interaction between the response latency and users’ subjective search experience in the mobile setting has been missing so far. To address this gap, we conduct a controlled study that aims to reveal how response latency affects mobile web search. Our preliminary results indicate that mobile web search users are four times more tolerant to response latency reported for desktop web search users. However, when exceeding a certain threshold of 7-10 sec, the delays have a sizeable impact and users report feeling significantly more \emph{tensed}, \emph{tired}, \emph{terrible}, \emph{frustrated} and \emph{sluggish}, all which contribute to a worse subjective user experience.

\end{abstract}



\maketitle

\section{Background and Motivation}
\label{sec:background}

Site performance has been, from an economic standpoint, a major determinant to a website's success~\cite{DAVIS2001249,DELLAERT199941,DENNIS2006810,Galletta2004,Jacko00,Fiona04,Taylor13,RAMSAY199877}. That is, a website that is characterised by relatively high page load times will not only degrade the experience of its users that are accustomed to sub-second response times, but will be also penalised with respect to its page ranking, since site speed is one of the factors of the ranking algorithm for many commercial search sites. 

In a joint conference presentation~\cite{google_bing}, Bing and Google scientists presented 
a series of latency experiments. Bing reported that an added latency of 1.5 seconds to their usual page loading times reduced the queries per user by 1.8\%, the revenues per user by 4.3\%, and the overall satisfaction by 3.8\%. In a similar vein, Google reported that a 400 ms delay resulted in a 0.59\% reduction in the number of searches per user and diminished traffic by 20\%. What's even more noteworthy, is that the slower user experience had a long-term effect on search behaviour Additionally, Amazon reported~\cite{amazon} that a 100 ms latency would result in a 1\% revenues drop, which amounts to a loss of \$745 million per year\footnote{Considering the company's estimated annual revenue of \$74.5 billion}. The aforementioned reports~\cite{google_bing, amazon} point out that a slow website can incur considerable negative effects on Search Engine Optimization (SEO), conversion rates, revenue, and user experience. 

Such findings have motivated a large body of research to investigate the response time of computer systems (refer to~\cite{DABROWSKI2011555} for an overview). In the more specific context of web systems, earlier work investigated the impact of page load time on web browsing behaviour~\cite{DAVIS2001249,RAMSAY199877,DELLAERT199941,DENNIS2006810,Galletta2004,Jacko00,Fiona04,Taylor13}. \citet{Taylor13} reported page load times tolerable by users who are seeking information online to be in the 7-11 seconds range. Despite being outdated,~\cite{Fiona04} also provides references to studies on identifying the largest page load time that users can tolerate. A related line of research has investigated the trade-off between the cost of searching in information seeking. However, the bulk of these studies were conducted over a decade ago, when people used primarily desktop computers to browse the internet and were accustomed to slower download speeds. Thus, it is not clear what exactly the \textit{expectations} of today's users are. 

More specifically, in~\cite{Maxwell14} the authors verified several hypotheses (taken from information foraging and search economic theories) about how users' search behaviour should change when faced with delays. \citet{Baskaya12} simulated interactive search sessions in a desktop scenario, where querying effort is low, and a smart phone scenario, which requires high querying effort. They showed that the user effort spent on searching, when coupled with a time constraint, determined the user search behaviour in both use cases. \citet{Schurman09} exposed the users of a commercial search site to varying response time delays and observed the impact on long-term search behaviour. \citet{Barreda-Angeles2015} conducted a controlled study to reveal the physiological effects of response latency on users and discovered that latency effects are present even at small increases in response latency, but this analysis is limited to the desktop setting. Last, \citet{Teevan13} performed a query log analysis on the impact of search latency on user engagement and found that, with increasing response latency, the likelihood that the user will click on the search results decreased.

Nowadays, web traffic is higher on mobile \cite{statcounter} and people use mostly mobile phones for their everyday tasks. In addition, 4G technology is an order of magnitude faster than the fixed lines of a decade ago. Thus, we need to assume that people have adapted to faster internet speeds, and that the expectations of what constitutes ``good'' internet may have risen. To this end, we revisit the work of understanding the effect of latency on user behaviour in the context of web search. Moreover, since previous work focused primarily on desktop search, 
we situate our analysis to the mobile setting and consider a number of simulated and increasing cellular network latency values, the effect of which is not well understood. 

\section{User Study}
\label{sec:user_study}

To demonstrate the impact of response latency on users' search behavior, we carried out a controlled experiment that examined users' interactions with different search sites. To avoid that participants became aware of the artificially-induced latency and that this awareness altered the results, we set-up our experiment in a way that the apparent goal was the subjective evaluation of search sites for answering complex tasks (unbeknownst to the participants, we manipulated the response latency for each question \& search site pair\footnote{The study was reviewed and approved by a team of legal experts.}). Additionally, to mitigate the potential effect of brand bias produced by well-known commercial search sites like Google, Bing or Yahoo Search~\cite{Bai:2017}, we opted for some less popular options.

\subsection{Design}
\label{ssec:design}

For our study, we used a three-way, mixed design (see \autoref{table:design}). The \emph{repeated measures} independent variables were as follows. First, the search site (with five levels: ``\texttt{\small DuckDuckGo}'', ``\texttt{\small Gibiru}'', ``\texttt{\small SearchEncrypt}'', ``\texttt{\small StartPage}'', ``\texttt{\small Yandex}'') was controlled by using one of the pre-selected commercial search sites to complete the task. Second, the response latency was controlled by pre-computing ten different latency values (measured in milliseconds: $l^{(3)}$:``337'', $l^{(4)}$:``506'', $l^{(5)}$:``759'', $l^{(6)}$:``1139'', $l^{(7)}$:``1709'', $l^{(8)}$:``2563'', $l^{(9)}$:``3844'',  $l^{(10)}$:``5767'', $l^{(11)}$:``8650'', $l^{(12)}$:``12975''). Considering that human perception is not linear but exponential (see Weber-Fechner Law \cite{vanderHelm2010,MacKay1213,staddon}), we increased the latency values super-linearly by applying the formula
\begin{equation}
    l^{(x)} = 1.5^{x} \times 100 \text{ (ms)}, x \in [3,12]
\end{equation}
The resulting values were then split into two sets of five levels each: \{$l^{(3)}_{1}$, $l^{(5)}_{2}$, $l^{(7)}_{3}$, $l^{(9)}_{4}$, $l^{(11)}_{5}$\} and \{$l^{(4)}_{1}$, $l^{(6)}_{2}$, $l^{(8)}_{3}$, $l^{(10)}_{4}$, $l^{(12)}_{5}$\}. Because of the relatively large number of trials, we allocated half of our participants to the first set and the other half to the second set. This way, we tested the effect of a wider range of response latencies and reduced the effort needed for completing the study. 

One may argue that such latency values are not informed by realistic time delays produced by any commercial search engine. This may apply to a desktop setting, where the main network latency components in a web search scenario are the network latency, search engine latency, and browser latency~\cite{Bai:2017}. However, mobile web browsing has different bottlenecks and resource constraints. Mobile devices are scaled down versions of desktops: they have limited CPU, poorer network capacities, and lower memory. Hence, it is not unlikely to observe latencies in the range of 5-10 seconds \cite{Butkiewicz15,Sivakumar14,Bocchi2016,Nejati16,Kelton17} due to the poor performance of the cellular network, the slower computational speeds, or other reasons, making it imperative to consider them in our analysis.

Also, considering the prior work on the effect of `anticipated time' on users' tolerance~\cite{bouch2000quality}, we introduced a \emph{between-groups} independent variable i.e. participants' prior expectations about the mobile network performance (two levels: ``anticipate connectivity issues'', ``expect usual internet speed''). To control this variable, we informed half of our participants that we previously experienced issues with the internet speed, so that they may expect connectivity issues. The dependent variable was the subjective user experience, as captured through self-reports.

{\def\arraystretch{1}
\begin{table}[!t]
{\footnotesize
\begin{subtable}{\linewidth}\centering
{\begin{tabular}{l|c|c} 
\toprule
\multicolumn{3}{c}{\textbf{Repeated measures}}\\
\midrule
Websites & \multicolumn{2}{c}{DuckDuckGo, Gibiru, SearchEncrypt, StartPage, Yandex} \\
\hline
Latency (ms)  & 338, 759, 1709, 3844, 8650 & 506, 1139, 2563, 5767, 12975
\\ 
\arrayrulecolor{black}\midrule
\multicolumn{3}{c}{\textbf{Between-group}}\\
\midrule
Expectation  & Anticipate connectivity issue & Expect usual internet speed
\\ 
\arrayrulecolor{black}\bottomrule
\end{tabular}}
\end{subtable}
}
\caption{Variables of Study Design.}
\label{table:design}
\vspace{-20pt}
\end{table}
}

\subsection{Apparatus}
\label{ssec:apparatus}

\subsubsection{Latency}
\label{sssec:latency}
We used a custom-made mobile phone app to access a starting web page that listed the five search sites used in our study. A hidden settings menu allowed the experimenter to set the artificially-induced latency before each search task. Following the definition of user-perceived latency proposed in~\cite{Bai:2017}, we considered as page load time the time difference between the rendering of the web page content in the user's browser and ``clicking`` on the corresponding link. To keep the latency values fixed, we used Android's WebView as browser and intercepted all the events where a link was clicked. We then made the browser view element invisible and replaced it with a text saying ``loading...''. The browser was made visible exactly after the time specified in the settings had expired. We conducted the experiment in a location with \textit{fast} and \textit{reliable} internet connection, so that the website would be fully loaded upon the web view becoming visible again. Each participant used their own mobile phone device.

\vspace*{-7pt}
\begin{table}[h!]
{\footnotesize
    \centering
    \caption{Example of a \emph{create} search task scenario.}
    \vspace{-1em}
    \begin{tabular}{| p{0.95\linewidth} |}
    \hline
    \textit{After the F1 season opened this year, your niece became really interested in soapbox derby racing. Since her parents are both really busy, you've agreed to help her build a F1 car so that she can enter a local race. The first step is to figure out how to build a car. Identify some basic designs that you might use and create a basic plan for constructing the car.} \\
    \hline
    \end{tabular}
    \label{table:example_search_task}
}
\end{table}
\vspace*{-9pt}

\subsubsection{Search Tasks}
\label{sssec:search_tasks}

Among a number of relevant query collections \cite{bailey2016uqv100,buckley1999trec}, we used the search tasks proposed by Kelly et al.~\cite{Kelly2015} that analyzes and categorizes the tasks by their complexity. The tasks follow Anderson and Krathwohl's taxonomy of educational objectives~\cite{Anderson1999}, which focuses on the cognitive process dimension and includes six types of cognitive processes: \emph{remember}, \emph{understand}, \emph{apply}, \emph{analyze}, \emph{evaluate} and \emph{create} (with increasing amounts of cognitive effort). Furthermore, it spans across the domains of health, commerce, entertainment, and science and technology. We opted only for those tasks that fall under the \emph{create} category (Table~\ref{table:example_search_task}), which require the searcher to find and compile a list of items, understand their differences and make a recommendation at the end of the search task. The reason for that is because the \emph{create} tasks were shown~\cite{Kelly2015} to result in significantly more search queries ($M = 4.85, SD = 4.42$), URLs visited ($M = 14.43, SD = 12.34$) and SERP clicks ($M = 5.98, SD = 5.02$), compared the other five types. Hence, they were more likely to facilitate sufficient exposure to the latency stimuli and observe its effects on the subjective user experience. Finally, we made minor adjustments to the content of the search tasks to keep it regionally relevant (e.g., using the term ``F1'' instead of ``NASCAR'' or ``football'' instead of ``baseball'').

\subsubsection{Experience Sampling}
\label{sssec:experience_sampling}
To quantify the impact of response latency on our participants, we administered a questionnaire at post-task. At first, we asked them to rate to what extent they were satisfied with the search results. This item served as a distractor, to make it more credible that we were studying the effectiveness of the search sites. Next, we prompted the participants to report their feedback on eight Likert-scales that made sense within the context of search sites: 1) three that inquired about the user's affective state, and 2) five from the Questionnaire for User Interaction Satisfaction (QUIS;~\cite{quis}). More specifically, we asked participants to rate their subjective experience with the search sites on the axes ``Bad--Good``, ``Tense--Calm'', ``Tired--Awake'', ``Terrible--Wonderful'', ``Difficult--Easy``, ``Frustrating--Satisfying'', ``Dull--Stimulating'', and ``Rigid--Flexible''. Each of these items used a 7-point Likert scale, where the center response represents the neutral point. 

\subsection{Participants}
\label{ssec:participants}
There were 30 participants (female=14, male=16), aged from 24 to 45 ($M=35.7, SD=5.3$) and free from any obvious physical or sensory impairment. The participants were of mixed nationality and were all proficient with the English language.

\subsection{Procedure}
\label{ssec:procedures}
At the beginning of the study, the participants were asked to fill out a demographics questionnaire. Then, they were told that the purpose of the experiment was to assess the utility of several search sites for answering complex tasks (without revealing its true purpose). To this end, they would have to evaluate to what extent the search sites allowed them to arrive to a satisfying answer, using their own mobile phone devices. The study consisted of five brief, informational search tasks (Section~\ref{sssec:search_tasks}), where participants were presented with a predefined scenario and were asked to freely issue as many search queries they wanted, and examine as many search results they needed, to address the search task objective. To control for order effects, the search task assignment and the sequence of artificially-induced latencies were randomized and then altered via the Latin-square method. This ensured that the search task or the search site would not have a systematic effect on the outcome, when grouping the results by the response latency values. 

To respect participants' time, we limited the time to answer each question to 5 minutes. We asked the participants to consider only websites and ignore video results, to avoid spending large parts of the allocated time in watching videos, as this would limit their exposure to the response latencies. We informed participants that they might not always arrive to a satisfying conclusion during that period of time. We also encouraged them to keep validating their results in case they arrived to a satisfying answer sooner. The rationale was to keep the exposure to the experimental manipulation comparable across all sessions. At the end of each search task, the participants were informed about the true conditions of the study and were asked to fill out a post-task questionnaire (Section~\ref{sssec:experience_sampling})

\begin{figure}[ht]
    \centering
    \includegraphics[width=0.64\linewidth, trim=0mm 1.2mm 0mm 0mm]{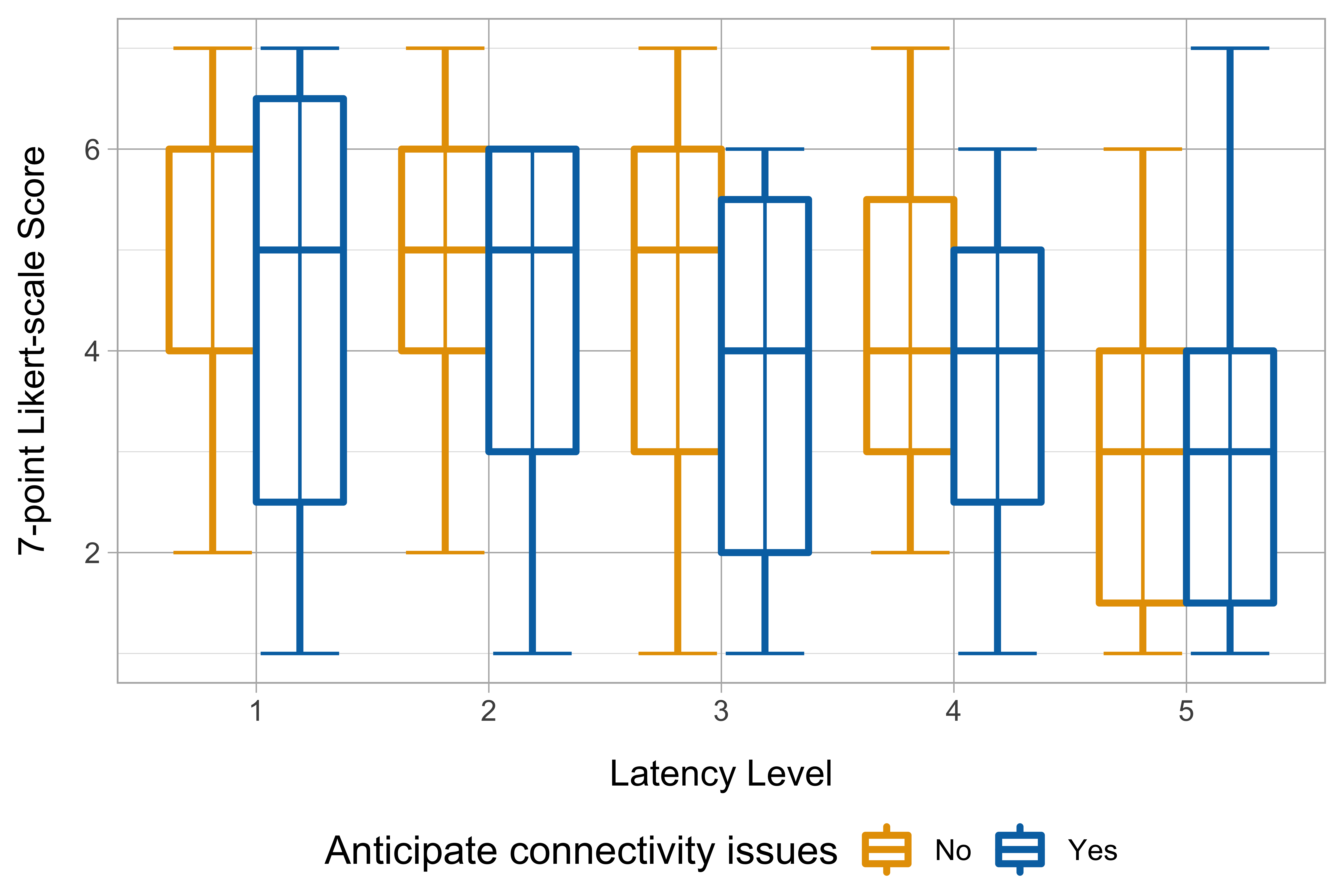}
    \vspace{-10pt}
    \caption{Boxplot of the `Frustrating vs. Satisfactory' Likert-scale scores ($1-7$) depending on the latency level.\label{fig:box}}
\vspace{-5pt}
\end{figure}

\section{Results \& Discussion}
\label{sec:results}
We employ two types of analysis. First, we use Friedman test\footnote{Friedman test was used as the ratings include repeated measures and their distribution did not seem normal.} as an omnibus test to detect the significant differences in the ratings among the response latency conditions. We then model the relationship between the ratings and response latency using ordinal regression. We elaborate on each analysis below. 

The Friedman test was combined with the Games Howell post-hoc test (with a Bonferroni correction) in order to conduct a pair-wise comparison of the ratings for all response latency levels. Prior to that, we tested and confirmed that our experimental control of the search site (e.g., ``\texttt{\small DuckDuckGo}``, ``\texttt{\small Gibiru}``) did not have any adverse effect on participants' ratings. A significant effect of the latency levels was identified for 5 out of the 8 subjective dimensions that we investigated: the affective dimensions ``Tense--Calm'' ($\chi^2_{(4)} = 7.600, p < .0001$) and ``Tired--Awake'' ($\chi^2_{(4)} = 12.90, p < .01$) and the QUIS items  ``Terrible--Wonderful'' ($\chi^2_{(4)} = 10.23, p < .05$), ``Frustrating--Satisfying'' ($\chi^2_{(4)} = 14.82, p < .01$) and ``Dull--Stimulating'' ($\chi^2_{(4)} = 13.36, p < .01$). Our post-hoc analysis suggests that users demonstrate a considerable amount of tolerance to response latency. For all five subjective dimensions, we did not observe any significant differences between response latency levels $l_{1}$ (338 ms, 506 ms) to $l_{4}$ (3844 ms, 5767 ms). However, the self-reported ratings were significantly lower for the highest response latency level $l_{5}$ (8650 ms, 12975 ms). For example, when examining the ``Frustrating--Satisfying'' scale (Fig.~\ref{fig:box}), the reported ratings for response latency $l_{5}$ dropped significantly compared to those of $l_{1}$, $l_{2}$ and $l_{3}$.

\begin{figure*}[!ht]
    \centering
    \includegraphics[width=0.94\textwidth, trim=0mm 1.2mm 0mm 0mm]{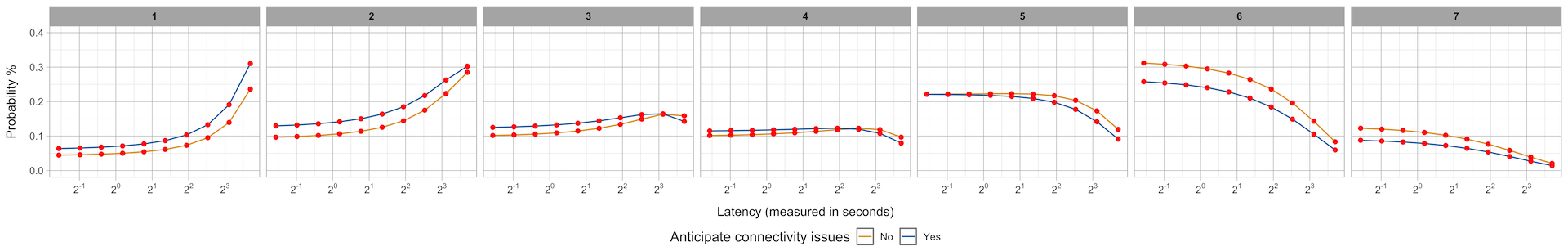}
    \caption{Estimated probability of the `Frustrating vs. Satisfactory' Likert-scale scores ($1-7$) depending on the latency. The red markers represent the response latency values (log2 scale) examined in this study.\label{fig:rating_prob}}
\end{figure*}

Similar to~\cite{Barreda-Angeles2015}, our self-report findings indicate
that users' ability to consciously perceive response latency degrades as we progress towards smaller delays. Interestingly, for the mobile setting, the user tolerance threshold for latency appears to be four times greater (7-10 sec) to that reported in~\cite{Barreda-Angeles2015, Teevan13}. We presume that the observed (high) tolerance could be related to factors such as keyboard layout, typing speed, and greater query formation effort, and may have a role in adjusting users' expectations~\cite{Baskaya12}. However, when exceeding that threshold, users felt significantly more \emph{tensed}, \emph{tired}, \emph{terrible}, \emph{frustrated} and \emph{sluggish}, all which contribute to a worse search experience. In addition, we did not find any significant differences on participants' ratings between the search sites or the prior expectations we set about the mobile network performance. We note that we increased the response latency values exponentially to simulate more accurately human perception of time. In practice, this means that the observed rating scores do not correspond to equally distanced response latency values, as done in~\cite{Barreda-Angeles2015, Teevan13}.

Using ordinal regression\footnote{We used the R package texttt{clmm2}~\cite{christensen2015tutorial}.}, we move beyond identifying the tolerance threshold and model how the ratings change as response latency increases along the continuous time scale i.e. we estimate the likelihood of each Likert-scale rating as a function of the response latency. For every scale in our questionnaire we run the model by taking the response latency and the prior expectation as the independent variables. As we have repeated measures, the random effect of the subjects is also accounted for in the model. 

{\def\arraystretch{0.6}
\begin{table}[ht]
{\footnotesize
\begin{subtable}{\linewidth}\centering
{\begin{tabular}{lcccccc} 
\toprule
Variable & Estimate & Std. Err. & z-value & Pr(>|z|) & Var. & Std. Dev. \\ 
\midrule
\multicolumn{7}{c}{\textbf{`Bad vs. Good'}}\\
\midrule
Latency & -0.1284 & 0.0389 & -3.2977 & 0.0009 & & \\ 
Expectation  & -0.2927 & 0.3065 & -0.9551 & 0.3395 & &  \\ 
\arrayrulecolor{gray!50!}\midrule
Rand. Effects & & & & & 0.0626 & 0.2502 \\
\arrayrulecolor{black}\midrule
\multicolumn{7}{c}{\textbf{`Difficult vs. Easy'}}\\
\midrule
Latency  & -0.1192 & 0.0389 & -3.0633 & 0.0002 & & \\ 
Expectation  & -0.3378 & 0.4282 & -0.9551 & 0.4302 & & \\ 
\arrayrulecolor{gray!50!}\midrule
Rand. Effects & & & & & 0.7140 & 0.8450 \\
\arrayrulecolor{black}\midrule
\multicolumn{7}{c}{\textbf{`Frustrating vs. Satisfying'}}\\
\midrule
Latency  & -0.1580 & 0.0400 & -3.9488 & 0.0001 & & \\ 
Expectation  & -0.3935 & 0.3232 & -1.2176 & 0.2234 & & \\ 
\arrayrulecolor{gray!50!}\midrule
Rand. Effects & & & & & 0.1338 & 0.3658 \\
\arrayrulecolor{black}\bottomrule
\end{tabular}}
\end{subtable}
}
\caption{Summary of ordinal regression analysis.}\label{table:ord_reg}
\vspace{-20pt}
\end{table}
}

For all the scales, the regression model confirms a very gradual decrease of the likelihood of observing positive ratings as the response latency increases, and a gradual increase for the negative ratings. Table \ref{table:ord_reg} shows the results for the scales that seem more relevant to web search tasks (the coefficients of the response latency for the remaining dimensions are significant and fall within the range of -.12 to -.15). The coefficient of -0.15 indicates that the odds of giving a lower rating (e.g., from rating 7 verses a rating below 7) is multiplied by 0.15 for every second of latency increase. Similar to our statistical analysis, the prior expectations about the mobile network performance do not seem to have a significant effect. In Fig.~\ref{fig:rating_prob} we transform the odds ratio to probabilities of each rating and plot the tendency along the time scale, for the scale ``Frustrating vs. Satisfying''. For example, the 6th column of Fig. \ref{fig:rating_prob} shows that the difference of the likelihood of rating 6 between the smallest and the largest latency is only about 20\%. Such gradual tendencies of the ratings demonstrate that users are less sensitive to small increases of latency, which is consistent to our previous finding. However, we also note that the likelihood of rating 1 (most negative) grows superlinearly (e.g., while the difference of the likelihood between 2'' and 4'' is around 2\%, that between 10'' and 12'' is around 5\%), implying that users' tolerance would quickly drop at a certain response latency limit.

\section{Conclusions}
\label{sec:conclusions}
Nowadays, web traffic is higher on mobile and people use mostly mobile phones for their everyday tasks. However, the page load performance on mobile devices does not match up to its importance: mobile page load times have been found to be an order of magnitude slower \cite{Butkiewicz15,Sivakumar14,Bocchi2016,Nejati16,Kelton17} compared to loading pages on a desktop browser, often taking 10s of seconds to load the landing page. With this in mind, we implemented a study where we examined the interaction between response latency and users’ subjective search experience in the mobile setting, and revisited previous work in light of today's adjusted expectations. 

Interestingly, our analysis indicates that users are four times more tolerant to known thresholds (1139-1709 ms) reported for the desktop setting, although the two cohorts of studies are not directly comparable. Through modelling the relationship between response latency and web search experience, we also show that the latter degrades quickly when the threshold of 7-10 sec is exceeded, which may explain the high \% of churn and revenue loss reported by several commercial search sites. This finding has several implications for web sites and browser vendors. First, it suggests that optimizations may not improve mobile and desktop web search equally, due to differences in users' expectations. Second, this tolerance zone creates an opportunity for mobile browsing: web page content may be served to each user at adjusted latencies, provided that no degradation in the user experience is expected. This setup requires less hardware resources on the search engine side, but also compensates for the limited CPU, poorer network capacities, and lower memory of mobile devices. Also, unlike~\cite{bouch2000quality}, we did not observe similar effects due to the prior expectations about the mobile network performance, for our setting. 

Last, our work comes with certain limitations. For example, our analysis is bounded by the power of the self-report tools we used and, as shown in \citet{Barreda-Angeles2015}, we cannot entirely discount the possibility that there are sizeable unconscious effects. In addition, our study involved a relatively small sample; we reserve the replication of these findings with a larger sample for future work. Furthermore, limiting the search task time to five minutes may have introduced a confounding factor of stress, which might have affected user behaviour. Finally, our findings relate to the complexity level of our search tasks; we intend to investigate the effects of response latency for other types of search tasks, and also of simpler nature.

\newpage

\balance{}

\bibliographystyle{ACM-Reference-Format}
\bibliography{sample-base}

\end{document}